\documentclass[10pt,a4paper]{iopart}
\pdfoutput=1

\usepackage{epsfig}
\usepackage{amsmath}
\usepackage{amsthm}
\usepackage{amssymb}
\usepackage{psfrag}
\usepackage[T1]{fontenc}
\usepackage{graphicx}
\usepackage{natbib}

\newtheorem{define}{Definition}
\newtheorem{prop}[define]{Proposition}
\newtheorem{rem}[define]{Remark}

\newcommand{\eps}{\varepsilon}
\newcommand{\w}{\omega}

\newcommand{\pcite}[1]{(see \cite{#1})}

\begin{document}

\title[]{Econometrics as Sorcery}

\author{G.~Innocenti and D.~Materassi}
\address{Dipartimento di Sistemi e Informatica \\
Centro per lo Studio di Dinamiche Complesse - CSDC \\
Università di Firenze, \\
via di S. Marta 3, 50139 Firenze, Italy\\ tel: +39 055 4796 360.}

\begin{abstract}
The paper deals with the problem of identifying the internal dependencies and similarities
among a large number of random processes.
Linear models are considered to describe the relations among the time series
and the energy associated to the corresponding modeling error is the criterion adopted
to quantify their similarities.
Such an approach is interpreted in terms of graph theory suggesting a natural
way to group processes together when one provides the best model to explain the other.
Moreover, the clustering technique introduced in this paper will turn out to be the dynamical generalization of other multivariate procedures described in literature.
\end{abstract}

\section{Introduction}\label{sec:intro}

Deriving information from data is a crucial problem in science and it has been widely investigated in literature.
A large variety of contributions has been developed in many fields, such as engineering, physics, biology and economy, providing several methods and procedures which accomplish to different objectives \cite{blatt96,rava02,palmer84,mant99}.
In particular, in the study of complex systems, the comprehension of the internal connections, which define the hierarchical structure of the process, turns out to play a key role to fully understand its dynamics.
This is especially true in presence of a multivariate data set, because this kind of samples is usually the result of a process intrinsically organized into interconnected subsystems \cite{mardia79}.
Therefore, the recognition of the system structure is a critical step for the definition of a suitable model.
In particular, a clusterization problem can be solved to divide the source data set into interconnected homogeneous groups describing different subsystems \cite{anderson73}.
This approach deals with the search of similarities and relations inside the original samples, trying to catch their internal connections and providing a schematic representation of hierarchies.
Recently, new clustering techniques based on a correlation matrix have been proposed for the analysis of data sets made up by a large variety of time series \cite{eisen1998clu,tum07}.
However, these procedures are able to detect only the ``static'' relations among the samples, since they capture the similarities just at the current time \citep{mant95,stan99a,nrm07}.

In this paper, we propose a clustering technique based on a modeling approach.
Indeed, since the original time series are dynamically interconnected, we intend to derive their hierarchy in terms of mathematical laws, which provide a structured description of the internal mechanics.
To this aim, we settle the clustering problem into the framework of the system identification theory \citep{ljung99,ksh00}.
Hence, exploiting the modeling errors to quantify the similarities among the original signals, we realize a clustering technique, defined as the solution of a minimization problem.
Therefore, a modeling interpretation of the procedures based on the correlation matrix is first introduced.
In particular, they turn out to be a non optimal choice with respect to the modeling error.
Then, the approach is developed taking into account dynamic dependencies among the time series.
To this regard, the identification step is realized introducing the hypothesis of linear dynamic connections, represented by Single Input-Single Output (SISO) local models.
Moreover, since the clusters are internally organized by means of transfer functions, the final model can be interpreted as a dynamical network of interconnected systems and its structure as the related topology.

~\\
\noindent{\bf Notation:}\\
$E[\cdot]$: mean operator; \\
$R_{XY}(\tau)\doteq E[X(t)Y(t+\tau)]$: cross-covariance function of stationary processes;\\
$R_{X}(\tau)\doteq R_{XX}(\tau)$: autocovariance;\\
$\rho_{XY}\doteq\frac{R_{XY}}{\sqrt{R_X R_Y}}$: correlation index;\\
$\mathcal{Z}(\cdot)$: Zeta-transform of a signal;\\
$\Phi_{XY}(z)\doteq\mathcal{Z}(R_{XY}(\tau))$: cross-power spectral density;\\
$\Phi_{X}(z)\doteq\Phi_{XX}(z)$: power spectral density;\\
with abuse of notation, $\Phi_{X}(\w)=\Phi_{X}(e^{i\w})$; \\
$\lceil \cdot \rceil$ and $\lfloor \cdot \rfloor$:
			ceiling and floor function respectively;\\
$(\cdot)^*$: complex conjugate.

\section{A Modeling Perspective}
In \cite{mant95} a procedure to obtain a hierarchical
structure of a set of time series is proposed.
$N$ realizations of $N$ random processes $X_i$ are considered.
First, an estimation of the
correlation index $\rho_{ij}$ related to every couple
$(X_i$, $X_j)$ is computed, along with the associated distances
\pcite{mant99}
\begin{align}\label{eq:Mantdistance}
	d_{ij}\doteq\sqrt{2(1-\rho_{ij})}~.
\end{align}
Then, a graph is defined where every node represents a random
process and the arc linking two nodes is weigthed according to
(\ref{eq:Mantdistance}).
Eventually, the Minimum Spanning Tree (MST) is extracted by the graph.
This procedure has been successfully exploited to provide a
quantitative and topological analysis of time series, expecially in
the economic field (see \cite{mant99}, \cite{tum07} and
\cite{nrm07}).
It is worth considering that such a technique can be interpreted
in terms of a modeling procedure.
Consider the problem of describing a process $X_j$ 
by scaling another process $X_i$ with a suitable real constant
$\alpha_{ji}$.
Choosing
\begin{align}\label{eq:mantconstgain}
	\alpha_{ji}=\sqrt{\frac{E[X_j^2]}{E[X_i^2]}}=
		\sqrt{\frac{R_{Xj}}{R_Xi}}~,
\end{align}
we find that
\begin{align*}
	&E\left[ (X_j - \alpha_{ji} X_i)^2 \right]=
				~E[X_j^2]~d_{ij}^2.
\end{align*}
Hence, the distance (\ref{eq:Mantdistance}) can be interpreted
as the root of the mean square error, properly normalized by
the variance of $X_j$,
when the simple gain (\ref{eq:mantconstgain}) is used.
Such a normalization is necessary since we are interested
into capturing similar trends between the processes regardless
of their amplitudes. However, we remark that the choice of
(\ref{eq:mantconstgain}) can be considered arbitrary.
Conversely, we would like to evaluate the closeness of two processes
according to the information which can be inferred about
one of them assuming to know the other \cite{gra69}.
From this point of view, (\ref{eq:mantconstgain}) does not satisfy
any optimality criterion. Indeed, considering two
anticorrelated time series ($\rho_{ij}=-1$) it is possible to perfectly
reconstruct one from the other. Thus the information in the two
signals is the same, while their distance 
(\ref{eq:Mantdistance}) makes them the farest away.\\
Let us define
\begin{align}
	e_{ji}=X_j-\alpha_{ji}X_i,
\end{align}
then, it is possible to adopt the least squares criterion in order to 
evaluate the ``best'' constant $\alpha_{ji}$.
In this case, it is immediate to prove that the optimal choice is given by
\begin{align}\label{eq:optconstgain}
	\hat \alpha_{ji}=\frac{R_{X_j X_i}}{R_{X_i}}
\end{align}
and the relative quadratic error amounts to
\begin{align}\label{eq:optgaincost}
	E[e_{ji}^2]=R_{X_j}-\frac{R_{X_j X_i}^2}{R_{X_i}}
\end{align}
\pcite{ksh00}.
In order to obtain an adimensional quantity, we can normalize
(\ref{eq:optgaincost}) with respect to the power of $X_j$ and
define the binary function
\begin{align}\label{eq:mantdist}
	&d(X_i,X_j)\doteq\sqrt{\frac{E[e_{ji}^2]}{R_{X_j}}}=
		\sqrt{1-\rho_{X_i X_j}^2}~.
\end{align}
It is worth observing that (\ref{eq:mantdist}) is a distance
exactly as (\ref{eq:Mantdistance}).
\begin{prop}\label{pr:trinequality-static}
	The function $d(\cdot,\cdot)$, as defined in (\ref{eq:mantdist}),
	is a metric.
\end{prop}
\textit{Proof.} See the appendix. $\hfill \square$

In \cite{mant95}, the MST is extracted from the
graph, according to the weights (\ref{eq:Mantdistance}).
This is equivalent to define a hierarchical structure of 
the time series relying on the adoption
of linear gain models (\ref{eq:mantconstgain})
between the processes and considering the relative modeling error
as a distance function.\\
Substituting (\ref{eq:Mantdistance}) with
(\ref{eq:mantdist}), we are applying the same topological strategy,
but we are structuring the data according to the the 
best gain model in the sense of the least squares.
\begin{rem}
From a system theory point of view, it can be said that both the
approaches are ``static''.
Indeed, the models do not have a state, thus they do not have any
dynamics.
They simply capture a direct relation between two process samples
at the same time instant.
However, the optimal approach we have followed can be
extended to a more general case.
\end{rem}

\section{Dynamic Modeling using Wiener Filters}\label{sec:noncausal}
We consider a model to be ``static'' (or ``memoryless'') when,
at every time instant $t$, its output is a function of its input at the very same time
instant.
Conversely, the output of a ``dynamic'' model also depends on the input values it receives
at instants different from $t$. In this general sense, we say that it has
a ``memory'' (or, equivalently, a ``state'').
Constant gains as (\ref{eq:mantconstgain}) or (\ref{eq:optconstgain})
are linear static models offering an extremely simple proportional
relation between two processes.
We propose a dynamic extension of the linear approach just described in the previous
section based on the well-known Wiener Filter.

Given two stochastic processes $X_i$, $X_j$
and a time discrete transfer function $W_{ji}(z)$ (that is the zeta-transform of its impulse  response),
let us consider the quadratic cost
\begin{align}\label{eq:cost}
	E\left[ (\eps_Q)^2 \right]
\end{align}
where 
\begin{align}
	\eps_Q\doteq Q(z)(X_j - W_{ji}(z)X_i)
\end{align}
being $Q(z)$ an arbitrary stable and causally invertible
time-discrete transfer function weighting the error 
\begin{align}
	e_{ji}=X_j - W_{ji}(z)X_i.
\end{align}
Then, the problem of evaluating the transfer function $\hat W(z)$
such that the quadratic cost (\ref{eq:cost}) is minimized 
is well-known in scientific literature and its
solution is referred to as the Wiener filter \pcite{ksh00}.
\begin{prop}[Wiener filter]
	The Wiener filter modeling $X_j$ by $X_i$ is the linear stable
	filter $\hat W_{ji}$ minimizing the filtered quantity
	(\ref{eq:cost}).
	Its expression is given by
	\begin{align}\label{eq:noncausalwiener}
		\hat W_{ji}(z) = \frac{\Phi_{X_i X_j}(z)}{\Phi_{X_i}(z)}
	\end{align}
	 and 	it does not depend upon $Q(z)$.
	Moreover, the minimized cost is equal to
	\begin{align*}
		\min E\left[(Q(z) \eps)^2\right]=
			\frac{1}{2\pi}\int_{-\pi}^{\pi} |Q(\w)|^2
			\left(\Phi_{X_j}(\omega)
			-|\Phi_{X_j X_i}(\w)|^2 \Phi_{X_i}^{-1}(\w) \right) d\w
	\end{align*}
\end{prop}
\textit{Proof.} See, for example, \cite{ksh00} 
$\hfill\square$

Observe that the stable implementation of the Wiener filter
$\hat W_{ji}(z)$ is non-causal, in general.
That is, its output $\hat W_{ji}(z)X_i$ depends
on both past and future values of the input process $X_i$.
The Wiener filter, in this formulation, is interesting from an
information and modeling point of view, but, of course,
we would rather need a causal filter, if we were to make predictions
(aim which is beyond the scope of this paper).\\
Since the weighting function $Q(z)$ does not affect the
Wiener filter, but only the energy of the filtered error,
we choose $Q(z)$ equal to $F_j(z)$, the inverse of the
spectral factor of $\Phi_{X_j}(z)$, that is
\begin{align}
	\Phi_{X_j}(z)=F_j^{-1}(z)(F_j^{-1}(z))^*
\end{align}
with $F_j(z)$ stable and causally invertible \pcite{kail01}.
In such a case the minimum cost assumes the value
\begin{align}\label{eq:coherentcost}
	\min E[\eps_{F_j}^2] = \frac{1}{2\pi}\int_{-\pi}^{\pi}
			\left(1- \frac{|\Phi_{X_j X_i}(\w)|^2}
				{\Phi_{X_i}(\w)\Phi_{X_j}(\w)}
			\right) d\w.
\end{align}
This peculiar choice of $Q(z)$ makes the cost depend explicitly
on the coherence function of the two processes
\begin{align}
	C_{X_iX_j}(\w)\doteq\frac{|\Phi_{X_jX_i}(\w)|^2}
					{\Phi_{X_i}(\w)\Phi_{X_j}(\w)}
\end{align}
which turns to be non negative and symmetric with respect to $\w$.
It is also well-known that the cross-spectral density satisfies the
Schwartz inequality. Hence, the coherence function is limited
between $0$ and $1$.
The choice $Q(z)=F_j(z)$ can be now understood as motivated by
the necessity to achieve an adimensional cost function not
depending on the power of the signals as in
(\ref{eq:coherentcost}).\\
The cost obtained by the minimization of the error $\eps_{F_j}$
using the Wiener filter as before allows us to define 
the binary function
\begin{align}\label{eq:distance}
	d(X_i,X_j)\doteq
			\left[ \frac{1}{2\pi}
				 \int_{-\pi}^{\pi}
				\left(1- 
					C_{X_{i} X_{j}}(\w)
				\right) d\w 
			\right]^{1/2}.
\end{align}
\begin{prop}\label{pr:trinequality-dynamic}
	The function $d(\cdot,\cdot)$, as defined in (\ref{eq:distance})
	is a metric.
\end{prop}
\textit{Proof.} See the appendix. $\hfill \square$

The metric (\ref{eq:distance}) can now be used to derive a MST
and obtain a hierarchical structure of the processes $X_i$.
Such an approach generalizes the results in \cite{mant95} to the
linear dynamic case.
We remark that the choice of a tree to describe the topology
of the data is a very reasonable but arbitrary solution.
In order to capture influences and similarities among the processes
$X_j$, we intend to propose a more flexible modeling technique
to extract topological information from the data.
Every $X_j$ can be described as the output of a linear SISO dynamical system,
whose input is one of the other $N-1$ processes.
Thus, for every time series $X_j$ it is natural to choose the model $\hat W_{j m(j)}$
with input $X_{m(j)}$, such that it provides the best description
according to (\ref{eq:coherentcost}), dropping the others.
The application of this procedure results in a set $N$ interconnected
systems, each of them minimizing $\min_{i} E[(Q_j e_{ji})^2]$.
Since the choice of every model $\hat W_{j m(j)}$ does not affect
the selection of the other ones, the overall cost function
\begin{align}\label{eq:generalcost}
	\min_{m(\cdot)} \sum_j E[(Q_j e_{jm(j)})^2]
\end{align}
turns out to be minimized, as well.
The following algorithm performs such a task.\\

\begin{figure}
	\begin{center}
		\includegraphics[width=0.55\columnwidth]{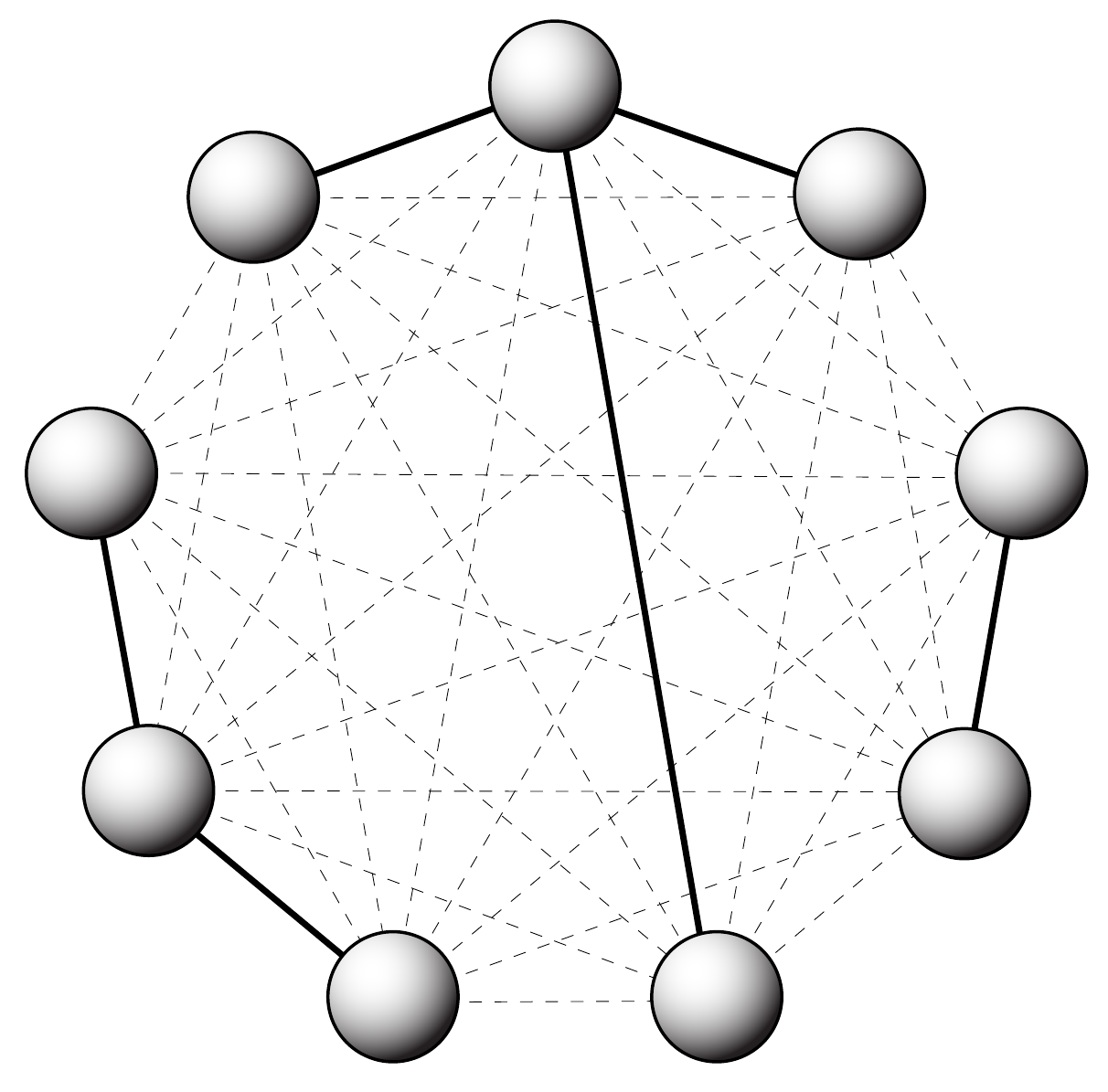}
		\caption{The figures illustrates 
				all the possible connections
				between two nodes (dashed lines)
				in a nine-node network.
				The solid lines depict a forest as it were
				the result after the application of the
				algoritm A.
			\label{fig:alga}}
	\end{center}
	\caption{The figure illustrates all the possible connections between two nodes
		(dashed lines) in a 10 nodes network. The solid lines depict a forest
		as it were the result after the application of the clusterization
		algorithm.}
\end{figure}

{\tt Clusterization Algorithm: }
\begin{itemize}
	\item[{\tt 1.}] initialize the set $A=\emptyset$
	\item[{\tt 2.}] for every process $X_j~ (j=1,...,N)$
	\begin{itemize}
		\item[{\tt 2a.}] for every  $i=1,...,N, i\neq j$\\
					compute the distance 
					$d_{ij}\doteq d(X_i,X_j)$;
		\item[{\tt 2b.}] define the set
				$M(j)\doteq\{ k| d_{kj}=\min_{i} d_{ij} \}$
		\item[{\tt 2c.}] choose, if possible, $m(j)\in M(j)$
			such that $(m(j),j) \notin A$
		\item[{\tt 2d.}] choose the model\\
			$X_j = \hat W_{jm(j)}(z) X_{m(j)} + e_{jm(j)}$
		\item[{\tt 2e.}] add the couple $(j,m(j))$ to $A$.
	\end{itemize}
\end{itemize}
The resulting network of processes has an appealing graphical interpretation.
Indeed, its topological structure can be
seen as a weigthed graph where every process $X_{j}$ is a node,
the arc linking $X_i$ to $X_j$ represents the Wiener Filter
describing the ``output'' $X_j$ in terms of the ``input'' $X_i$
and the weights on the arcs are given by (\ref{eq:distance}).
Because of the simmetry property of (\ref{eq:distance}), there
is no actual need to consider an oriented graph. Hence, the
presence of both the arcs $(i,j)$ and $(j,i)$ boils down into just
a single link.
Following this interpretation, the algorithm determinates a graph 
designed to keep, for every node, the incident arc with the least cost
(see Figure~\ref{fig:alga}).
\begin{prop}\label{pr:graphproperties-noncausal}
The graph resulting from the proposed algorithm has the following
properties:
\begin{itemize}
	\item on every node there is at least an incident arc
	\item if there is a cycle, then all the arcs of the cycle
			have the same weight
	\item there are at least $\lceil N/2 \rceil$
		and at most $N$ arcs.
\end{itemize}
\end{prop}
\textit{Proof.} See the appendix. $\hfill \square$

The presence of cycles in the resulting graph
is a pathological situation as stressed in the following remark.
\begin{rem}
	A necessary condition of existence for a cycle
	is the presence of more than two nodes with common multiple
	minimum cost arcs. Therefore, a mild sufficient condition
	in order to avoid cycles in the graph is to assume that
	every node has a unique minimum cost arc.
	If the costs of the arcs are obtained by estimation from
	real data the probability to obtain a cycle is zero almost
	everywhere \pcite{shir95}.
	Consequently, in such a case the expected
	topology of the graph is a forest (a graph with no cycles).
\end{rem}
\begin{rem}
If there are no cycles, the graph resulting from the algorithm is
a subgraph of the MST.
\end{rem}
\begin{rem}
In general, nothing can be said about the connectivity.
Therefore, the modeling procedure depicted by the
algorithm provides a clusterization of the original processes $X_i$
which, for every node, minimizes the cost (\ref{eq:distance})
according to the criterion of linear dynamic dependency.
It is possible to modify the procedure
in order to suitably satisfy other constraints about the graph
topology.
For instance, if we deal with a connectivity condition the 
algorithm can be easily replaced by a MST
search.
Therefore, the approach followed in \cite{mant95} to obtain
topological information from the time series results in a
constrained optimization of (\ref{eq:generalcost}).
\end{rem}
\begin{rem}
The modelization we have derived makes use of  non causal
Wiener filters, thus it can be useful to detect linear dependencies
of any sort between the processes $X_i$.
\end{rem}
Unfortunately, the adoption of non causal filters can not be
employed to make predictions.

\section{Numerical Example}
\begin{figure}
	\begin{center}
		\includegraphics[width=0.55\textwidth]{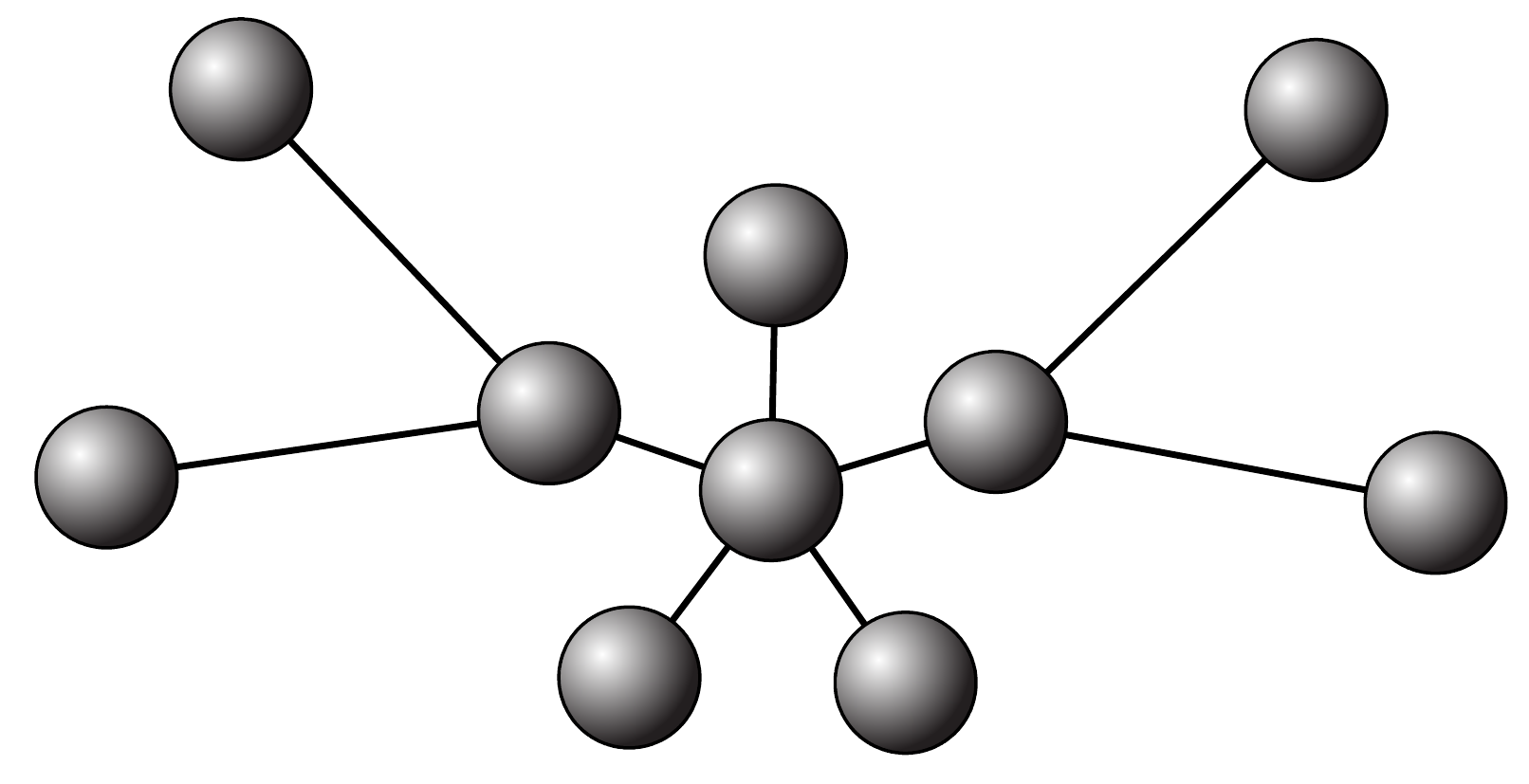}
	\end{center}
	\caption{The figure illustrates the topology of the 10 nodes network analyzed
		in the numerical examples paragraph.
		Each node represent a process $X_j$, while the arcs describe the
		connections among them, according to the linear SISO model \eqref{eq:simmodels}.
		For the data generation we have considered only transfer functions of
		at most the second order.
		The noises $N_j$ have been assumed to provide half the power of the
		affected processes.
		The samples have been collected over 1000 time steps.
		\label{fig:truetreeN10_002}}
\end{figure}

\begin{figure}
	\begin{center}
		\includegraphics[width=0.55\textwidth]{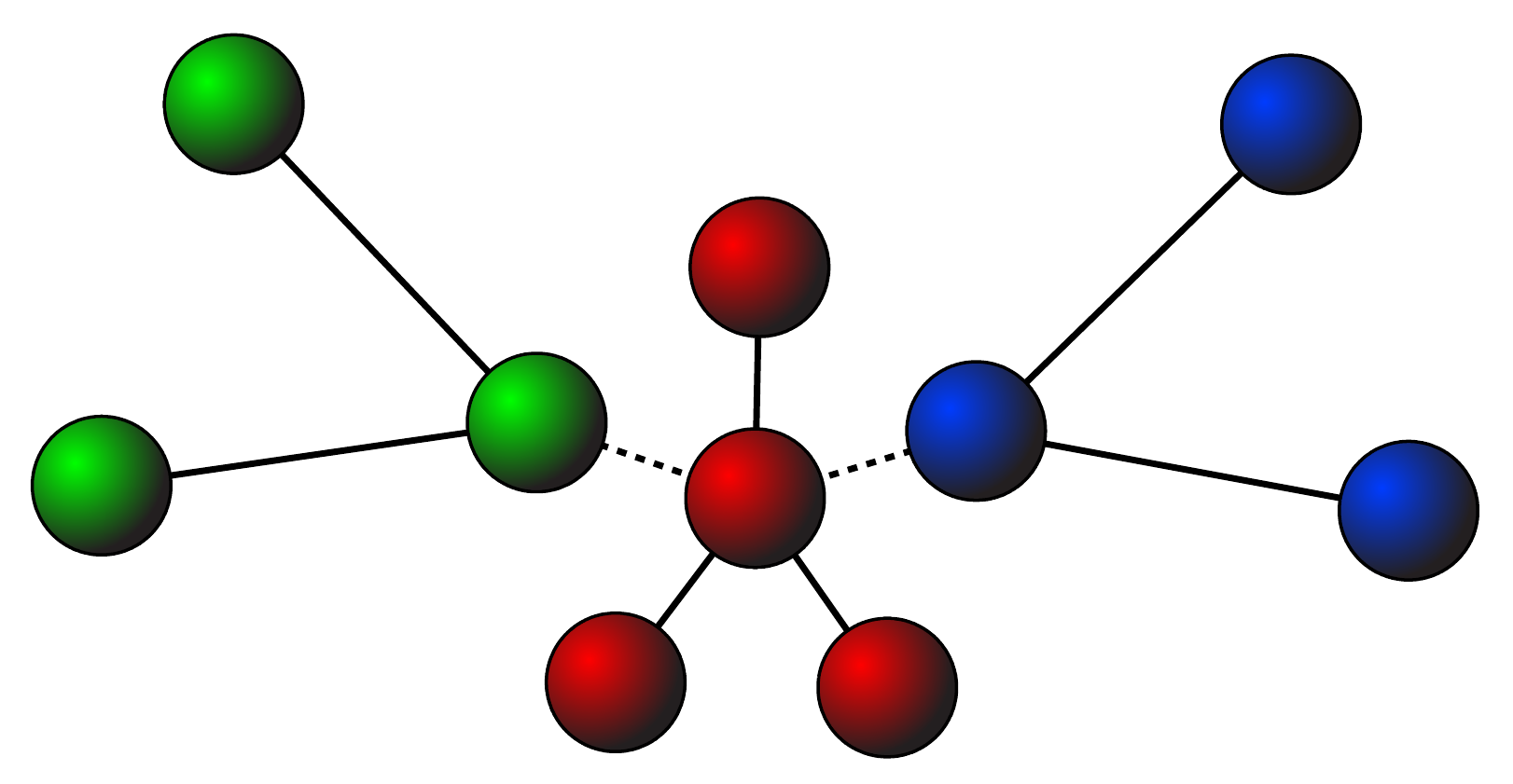}
	\end{center}
	\caption{The figure illustrates the MST obtained by using the coherence-based
		distance (solid+dashed lines).
		Notably, it is the same of the actual topology.
		The application of the proposed clustering algorithm provides a forest
		(solid lines): each cluster is virtually connected to the others by
		the arcs of the MST, which have not been chosen by the algorithm
		(dashed lines).
		The use of different colors (online) highlights the modular structure
		resulting from the clusterization.
		The presence of a very high noise-to-signal ratio prevents the
		algorithm to correctly reconstruct the actual topology, if no
		connectivity constraint is given.
		\label{fig:dgtreeN10_002}}
\end{figure}

\begin{figure}
	\begin{center}
		\includegraphics[width=0.55\textwidth]{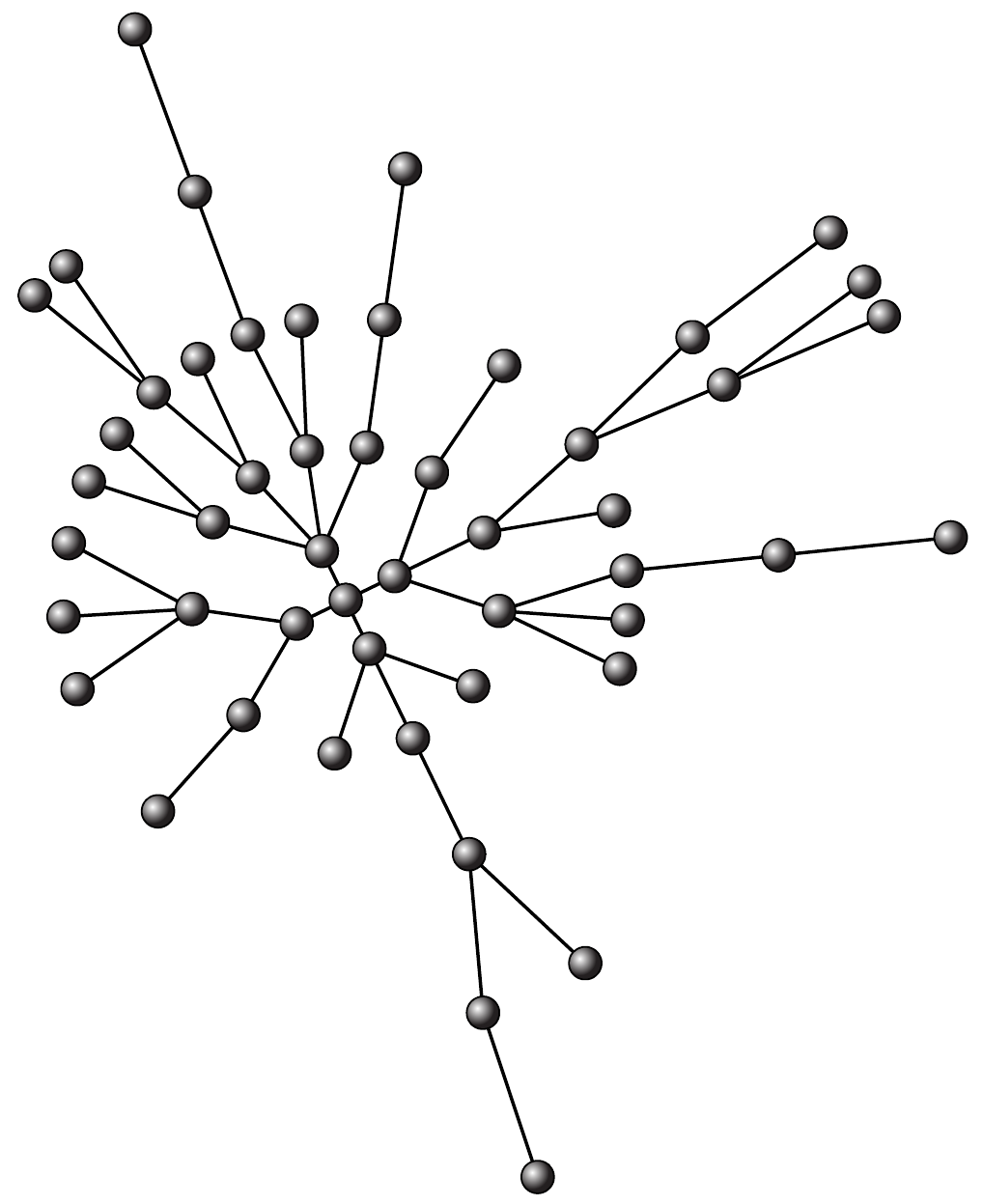}
	\end{center}
	\caption{
		The 50 nodes network of the numerical examples paragraph.
		The figure provides the actual topology.
		The example has been designed according to the same assumptions of the
		network of Figure \ref{fig:topology50processes}.
		\label{fig:topology50processes}}
\end{figure}

\begin{figure}
	\begin{center}
		\includegraphics[width=0.55\textwidth]{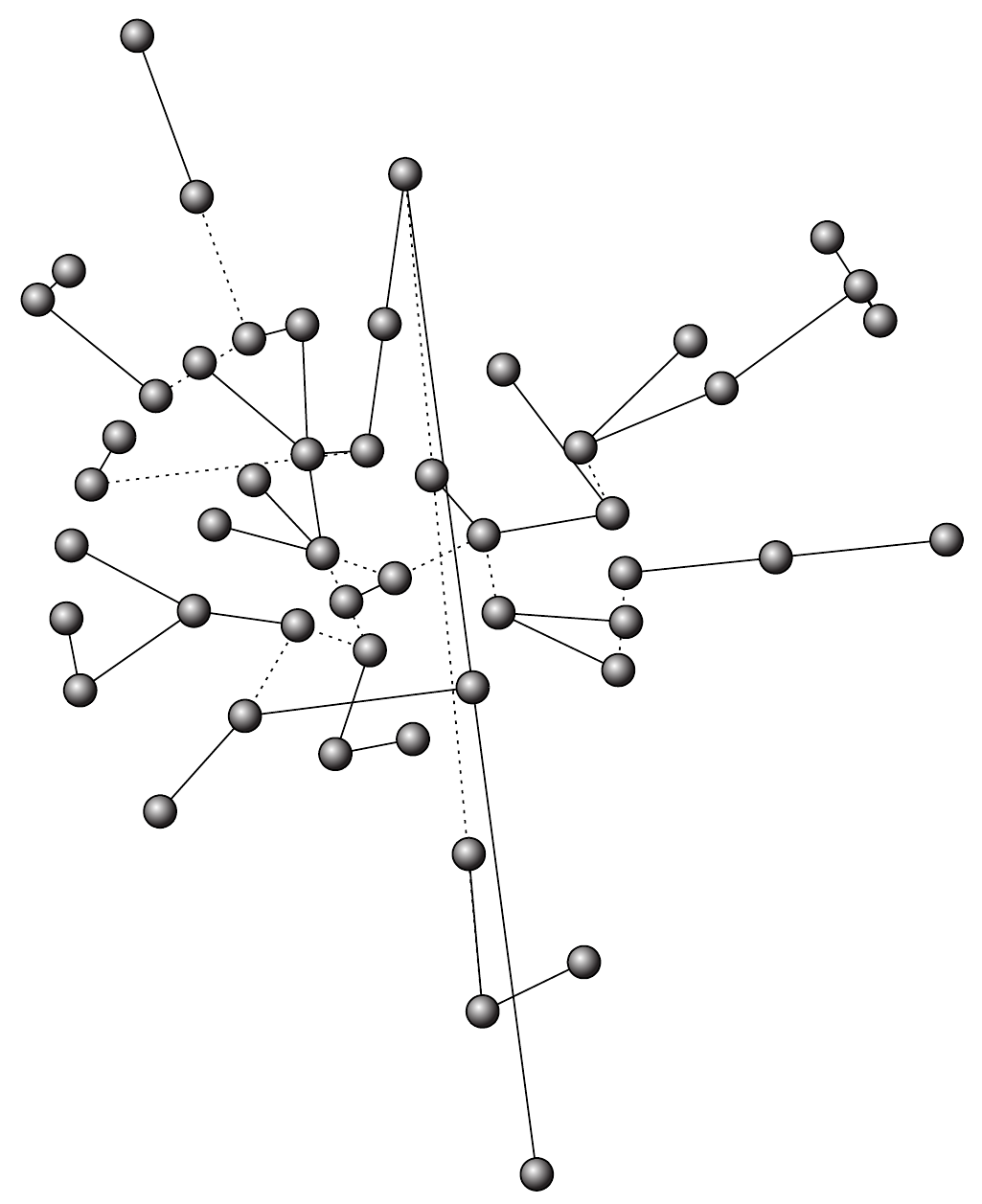}
	\end{center}
	\caption{The figure illustrates the MST obtaining by means of the
		correlation-based distance.
		Though the original topology is a tree, a quite significative amount of connections
		have not been correctly reconstructed.
		A limited number of similarities with the actual network can be observed.
\label{fig:manttree}}
\end{figure}

\begin{figure}
	\begin{center}
		\includegraphics[width=0.55\textwidth]{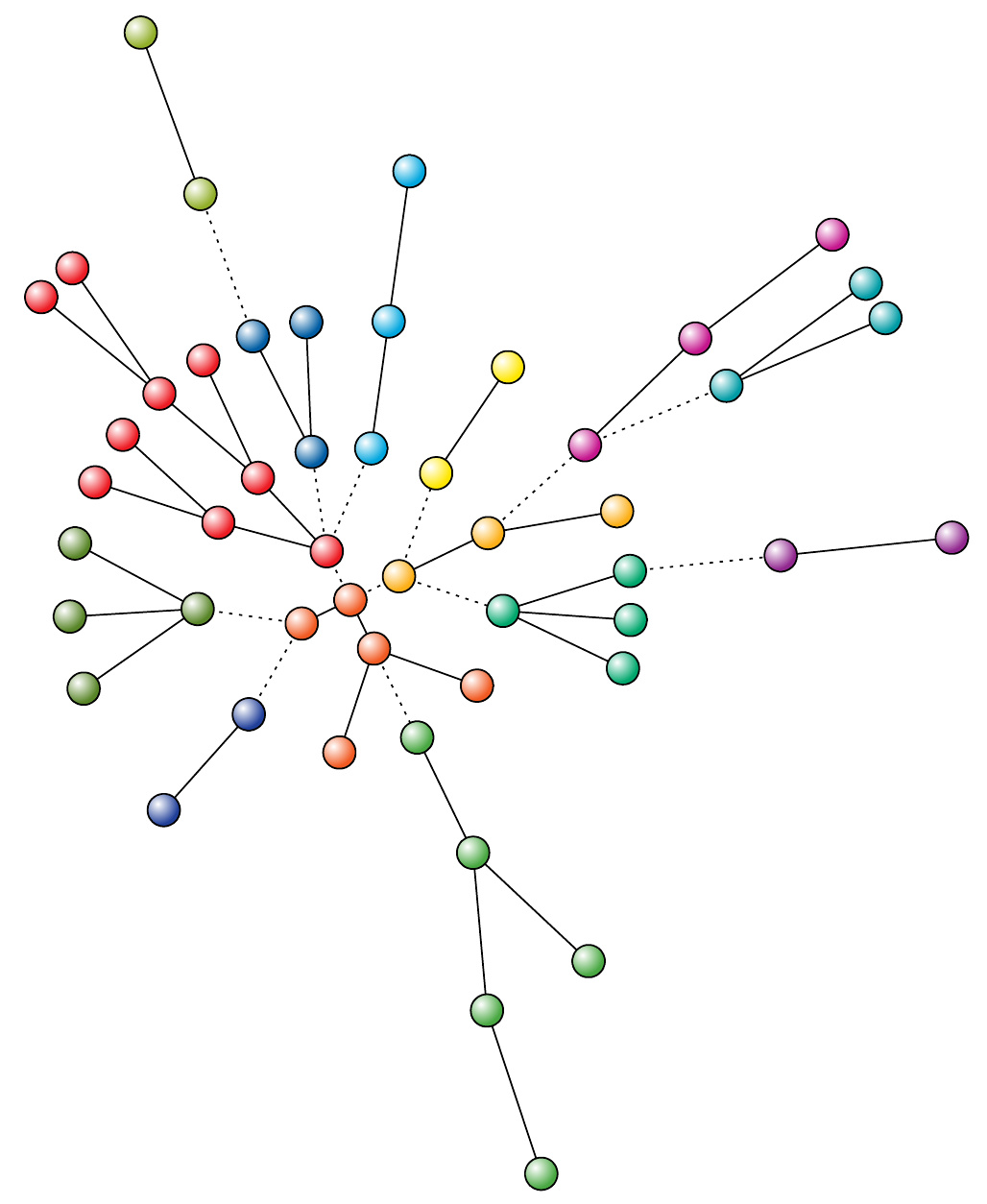}
	\end{center}
	\caption{The figure show the MST (solid+dashed lines) obtained by applying the
		coherence-based distance (\ref{eq:simmodels}) to the processes produced by the
		network depicted in Figure \ref{fig:topology50processes}.
		Notably, the original structure has been correctly reconstructed.
		The forest resulting from the application of the clusterization
		algorithm is also reported (solid lines).
		The clusters (colors online) result connected by the remaining arc of
		the MST (dashed lines).
		The algorithm does not manage to reconstruct exactly one cluster,
		because of the high noise-to-signal ratio.\label{fig:coherencetree}}
\end{figure}

It is intended to show, by means of numerical examples, the main advantages of the technique
described in the previous sections.
In particular, we want to evaluate the performance of our procedure
when identifying an unknown topology.
First, we realized several simulations of $10$ randomly generated processes $X_j$,
designed as follows.
They have been hierarchiacally structured in a tree topology, where the interconnections
were linear, randomly generated, at most second order transfer functions $W_{ji}$
with external noise $N_j$.
\begin{align}\label{eq:simmodels}
	X_{ji}= W_{ji}(z) X_i +N_j
\end{align}
Since all simulations present strong analogies, we are showing just one of them, whose
topology is depicted in Figure \ref{fig:truetreeN10_002}.
Note that the  simulated network involves linear dependencies only, so it satisfies the
theoretical conditions of the approach based on Wiener filters introduced in the previous
sections.
On every node $X_j$ (but the root) the deterministic component $W_{ji}(z) X_i$ and
the stochastic disturbance are equal in power.
A simulation horizon of $1000$ steps has been taken into account where the noise components
have been generated by pseudo random number algorithms. The incorrelation hypothesis
among the disturbances has been numerically checked providing a marginally satisfactory
result.
Applying the correlation technique described in \cite{mant99}, we found
the distance matrix reported in Table~\ref{tab:mantdistsim} and
the corresponding MST depicted in Figure \ref{fig:manttree}.
\begin{table}[ht]
	\centering		
	{\tiny
	\begin{tabular}{|c|cccccccccc|}
			\hline
		& $X_1$ & $X_2$ & $X_3$ & $X_4$ & $X_5$ & $X_6$ & $X_7$ & $X_8$ & $X_9$ & $X_{10}$ \\
		\hline
		$X_1$ & 0	& 0.9946	& 1.3763	& 1.0624	& 1.1027	& 1.2393	& 1.2719	& 1.3747	& 0.7306	& 1.4589 \\
		$X_2$ & 0.9946	& 3.3e-8	& 1.1130	& 1.0674	& 0.7723	& 1.0082	& 1.2004	& 1.1269	& 1.1132	& 1.4575 \\
		$X_3$ & 1.3763	& 1.1130	& 0	& 1.1487	& 1.2217	& 1.2877	& 1.1645	& 0.9965	& 1.3507	& 1.4124 \\
		$X_4$ & 1.0624	& 1.0674	& 1.1487	& 4.2e-8	& 1.1727	& 1.1805	& 0.9296	& 1.1455	& 1.1491	& 1.3433 \\
		$X_5$ & 1.1027	& 0.7723	& 1.2217	& 1.1727	& 3.9e-8	& 1.1491	& 1.2418	& 1.2353	& 1.1898	& 1.4587 \\
		$X_6$ & 1.2393	& 1.0082	& 1.2877	& 1.1805	& 1.1491	& 4.9e-8	& 1.2123	& 1.2984	& 1.2858	& 1.3227 \\
		$X_7$ & 1.2719	& 1.2004	& 1.1645	& 0.9296	& 1.2418	& 1.2123	& 0	& 1.1815	& 1.3003	& 1.3334 \\
		$X_8$ & 1.3747	& 1.1269	& 0.9965	& 1.1455	& 1.2353	& 1.2984	& 1.1815	& 0	& 1.3542	& 1.4389 \\
		$X_9$ & 0.7306	& 1.1132	& 1.3507	& 1.1491	& 1.1898	& 1.2858	& 1.3003	& 1.3542	& 7.3e-8	& 1.4450 \\
		$X_{10}$ & 1.4589	& 1.4575	& 1.4124	& 1.3433	& 1.4587	& 1.3227	& 1.3334	& 1.4389	& 1.4450	& 0 \\
		\hline
	\end{tabular}
	\caption{Correlation-based distance matrix. \label{tab:mantdistsim}}
	}
\end{table}
\begin{figure}
	\begin{center}
		\includegraphics[width=0.55\textwidth]{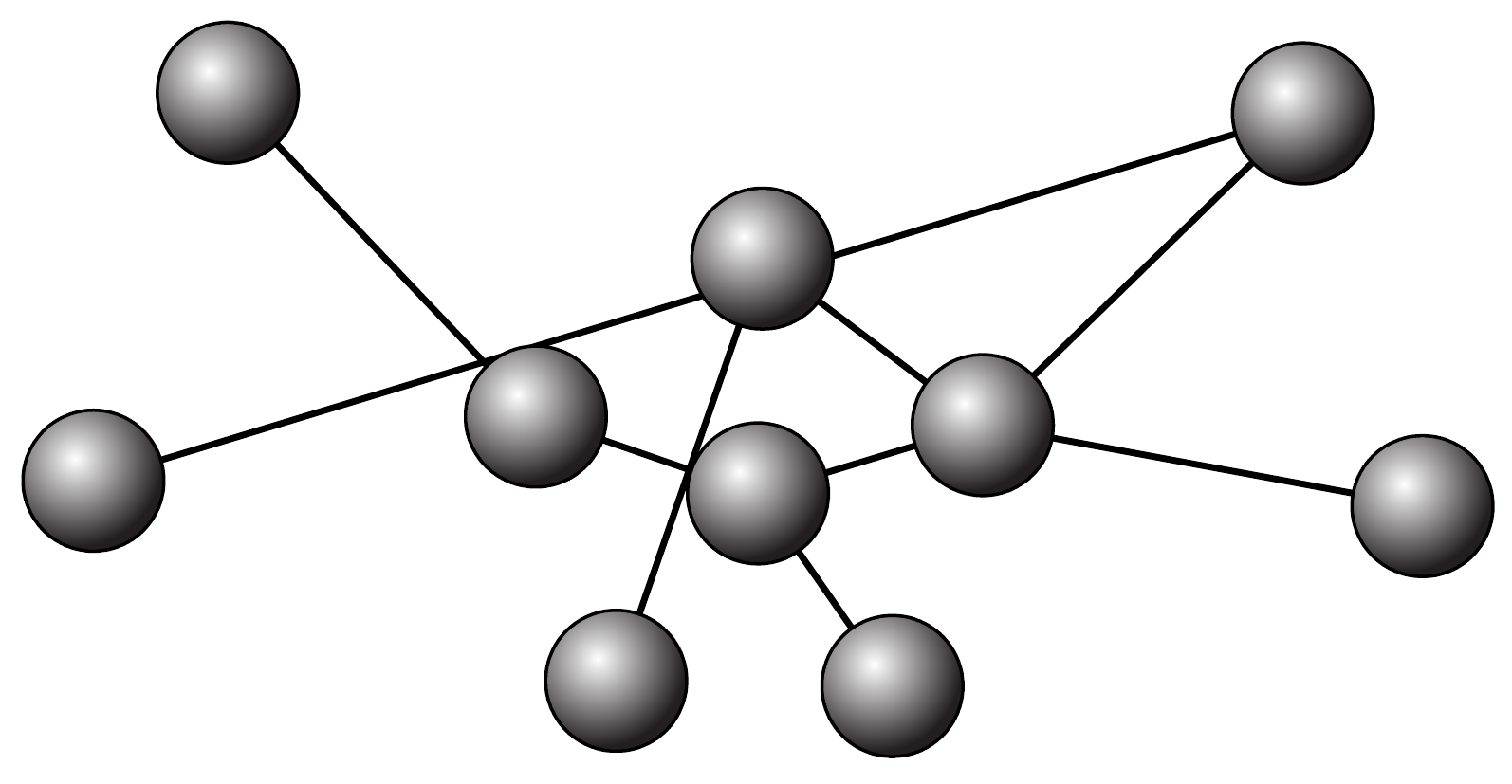}
	\end{center}
	\caption{The MST obtained using the correlation-based distance of Table(\ref{tab:mantdistsim}).
		The actual topology has not been correctly identified, though some
		analogies with the right structure can be observed.
		The procedure described in \cite{mant99} reveals strong limitation in
		capture the nature of the network even when the actual topology is
		exactly a tree.\label{fig:manttreeN10_002}}
\end{figure}
We note that the topology is not correctly identified by such a procedure, even though
similarities can be identified.
On the other hand, the application of the clusterization algorithm introduced by us
provides
the distances of Table~\ref{tab:coherencedistsim} with 
the graph of Figure~\ref{fig:dgtreeN10_002}.
\begin{table}[ht]
	{\tiny
	\centering
	\begin{tabular}{|c|cccccccccc|}
	\hline
		& $X_1$ & $X_2$ & $X_3$ & $X_4$ & $X_5$ & $X_6$ & $X_7$ & $X_8$ & $X_9$ & $X_{10}$ \\
		\hline
		$X_1$ & 0	& 0,7299	& 0,6675	& 0,7351	& 0,8316	& 0,8542	& 0,8297	& 0,7055	& 0,6549	& 0,8298 \\
		$X_2$ & 0,7299	& 0	& 0,8065	& 0,8353	& 0,6934	& 0,7358	& 0,8786	& 0,8483	& 0,8299	& 0,8717 \\
		$X_3$ & 0,6675	& 0,8065	& 0	& 0,8216	& 0,8744	& 0,8807	& 0,8750	& 0,8262	& 0,7841	& 0,8821 \\
		$X_4$ & 0,7351	& 0,8353	& 0,8216	& 0	& 0,8662	& 0,8722	& 0,7404	& 0,8502	& 0,8198	& 0,7039 \\
		$X_5$ & 0,8316	& 0,6934	& 0,8744	& 0,8662	& 0	& 0,8540	& 0,8919	& 0,8995	& 0,8730	& 0,8846 \\
		$X_6$ & 0,8542	& 0,7358	& 0,8807	& 0,8722	& 0,8540	& 0	& 0,8934	& 0,8984	& 0,8796	& 0,8944 \\
		$X_7$ & 0,8297	& 0,8786	& 0,8750	& 0,7404	& 0,8919	& 0,8934	& 0	& 0,8838	& 0,8694	& 0,8346 \\
		$X_8$ & 0,7055	& 0,8483	& 0,8262	& 0,8502	& 0,8995	& 0,8984	& 0,8838	& 0	& 0,8167	& 0,8908 \\
		$X_9$ & 0,6549	& 0,8299	& 0,7841	& 0,8198	& 0,8730	& 0,8796	& 0,8694	& 0,8167	& 0	& 0,8715 \\
		$X_{10}$ & 0,8298	& 0,8717	& 0,8821	& 0,7039	& 0,8846	& 0,8944	& 0,8346	& 0,8908	& 0,8715	& 0 \\
		\hline
	\end{tabular}
	\caption{Coherence-based distance matrix. \label{tab:coherencedistsim}}
	}
\end{table}
We stress that the topology is perfectly reconstructed by the procedure
if the connectivity constraint is imposed.
Further, we repeated the same procedure with a larger number of processes ($N=50$).
Again, results showed many similarities, so we are presenting just one case with
the topology depicted in Figure \ref{fig:topology50processes}
Analogously, the correlation-based approach provides the MST of Figure \ref{fig:manttree}
while the coherence-based algorithm identifies the graph of Figure \ref{fig:coherencetree}.
It is worth noting that our technique detects links actually present in the
topology. However the presence of a low signal to noise ration prevents
the complete reconstruction of the original tree topology in the absence
of any connectivity constraints.
These simple examples highlight a better capability of our technique
into capturing relationships and dependencies among time series.
In particular, remarkable improvements should be expected in
presence of strong dynamical interconnections and significative delays
in the actual network.

\section{Conclusions}\label{sec:conclusion}

In this paper we have introduced a novel approach to the clusterization problem.
In particular, the similarities among the time series of a
multivariate data set have been analyzed from the modeling point of
view and their interconnections have been interpreted as functional
dependencies.
Hence, linear SISO transfer functions have been proposed to describe
the relations among the processes and the associated modeling errors
have been exploited to quantify their similarities.
In turn, such a distance introduces a natural way of grouping the time
series, since it is very reasonable to place two processes in the same
subset, when one provides the best model to explain the other.
Notably, the proposed distance can be directly computed exploiting the
coherence function, requiring no identification step.
Further, our novel approach has been compared to the clustering
technique proposed in \cite{mant99} and formulated as an extension of the
multivariate analysis of \cite{tum07}.
In particular, our coherence-based distance turns out to be the
dynamical generalization of the correlation-based metric in \cite{mant95}.
Therefore, it provides an improved capability in capturing the
internal topology among the processes, especially when their
functional dependencies turns out to be dynamical laws.
Some numerical examples have finally been presented to illustrate the
expected improvements, due to our distance, and to provide a
validation for our clustering algorithm.

\section{Appendix}
\textit{Proof of Proposition \ref{pr:trinequality-static} .}
Note that we consider two processes be equivalent also when they are anticorrelated
since they are identical from an information point of view.
Thus, the only non trivial property to show is the triangle inequality.
Consider the following relations involving the optimal gains
$\hat \alpha_{31}$, $\hat \alpha_{32}$, $\hat \alpha_{21}$
\begin{align*}
	&X_3=\hat \alpha_{31} X_1 + e_{31}\\
	&X_3=\hat \alpha_{32} X_2 + e_{32}\\
	&X_2=\hat \alpha_{21} X_1 + e_{21}~.
\end{align*}
Since $\hat \alpha_{31}$ is the best constant model, we have that
it must perform better than any other constant model (in
particular $\hat \alpha_{32} \hat \alpha_{21}$)
\begin{align*}
	R_{X_3}-\frac{R_{X_3 X_1}^2}{R_{X_1}}\leq
		E[(e_{32} + \hat \alpha_{32} e_{21})^2]\leq
		\left(\sqrt{E[e_{32}^2]} +
		|\hat \alpha_{32}| \sqrt{E[e_{21})^2]} \right)^2~.
\end{align*}
Normalize with respect to $R_{X_3}$ and consider the square root
\begin{align*}
	\sqrt{1-\rho^2_{X_1 X_3}}&\leq 
		\sqrt{\frac{1}{R_{X_3}} \left(\sqrt{E[e_{32}^2]} +
		|\hat \alpha_{32}| \sqrt{E[e_{21}^2]} \right)^2}\leq\\
		&=\sqrt{\frac{E[e_{32}^2]}{R_{X_3}}}+
		|\rho_{X_2 X_3}|\sqrt{\frac{E[e_{21}^2]}{R_{X_2}}}.
\end{align*}
Since $|\rho_{X_2 X_3}|\leq 1$, we have the assertion.
$\hfill \square$

\textit{Proof of Proposition \ref{pr:trinequality-dynamic}} 
The only non trivial property to prove is the triangle inequality.
Let $\hat W_{ji}(z)$ be the Wiener filter between
$X_i, X_j$ computed according
to (\ref{eq:noncausalwiener}) and $e_{ji}$ the relative error.
The following relations hold:
\begin{align*}
	&X_3=\hat W_{31}(z) X_1 + e_{31}\\
	&X_3=\hat W_{32}(z) X_2 + e_{32}\\
	&X_2=\hat W_{21}(z) X_1 + e_{21}.
\end{align*}
Since $\hat W_{31}(z)$ is the Wiener filter between the two
processes $X_1$ and $X_3$, it performs better at any frequency
than any other linear filter, such as
$\hat W_{32}(z) \hat W_{21}(z)$.
So we have\\
\begin{align*}
	\Phi_{e_{31}}(\w)
		&\leq \Phi_{e_{32}}(\w) +
		|\hat W_{32}(\w)|^2 \Phi_{e_{21}}(\w) +\\
		&+\Phi_{e_{32} e_{21}}(\w)  \hat W_{32}^*(\w)+
		\hat W_{32}(\w) \Phi_{e_{21} e_{32}}(\w) \leq\\
		&\leq(\sqrt{\Phi_{e_{32}}(\w)}
			 + |\hat W_{32}(\w)|\sqrt{\Phi_{e_{21}}(\w)})^2
		\quad \forall~\w\in\mathbb{R}.
\end{align*}
For the sake of simplicity we neglect to explicitly write
the argument $\w$ in the following passages. 
Normalizing with respect to $\Phi_{X_{3}}$, we find
\begin{align*}
	\frac{\Phi_{e_{31}}}{\Phi_{X_3}}\leq
		\frac{1}{\Phi_{X_{3}}}
		(\sqrt{\Phi_{e_{32}}} +
			|\hat W_{32}|\sqrt { \Phi_{e_{21}} })^2
\end{align*}
and considering the 2-norm properties
\begin{align*}
	\left( \int_{-\pi}^{\pi}
		\frac{\Phi_{e_{31}}}{\Phi_{X_3}}d\w\right)^\frac{1}{2}
		\leq\left( \int_{-\pi}^{\pi}
		\frac{\Phi_{e_{32}}}{\Phi_{X_{3}}}d\w
		\right)^\frac{1}{2}
		+\left( \int_{-\pi}^{\pi}
			\frac{|\Phi_{X_3 X_2}|^2}{ \Phi_{X_{3}} \Phi_{X_{2}}}
			\frac{ \Phi_{e_{21}} } {\Phi_{X_{2}}}d\w
		\right)^\frac{1}{2}
\end{align*}
where we have substituted the expression of $\hat W_{32}$.
Finally, considering that
\begin{align*}
	0\leq \frac{|\Phi_{X_3 X_2}|^2}{ \Phi_{X_{3}} \Phi_{X_{2}}}
		\leq 1,
\end{align*}
we find
\begin{align*}
	d(X_1,X_3) \leq d(X_1,X_2) + d(X_2,X_3).
\end{align*}
$\hfill\square$

\textit{Proof of Proposition \ref{pr:graphproperties-noncausal}}
The proof of the first property is straightforward because for every
node the algorithm considers an incident arc.
Let us suppose there is a cycle and be $k$ the number of nodes
$n_1,...,n_k$ and arcs $a_1,...,a_k$ of such a cycle.
Every arc $a_1,...,a_k$ has been chosen at the step
{\tt 2e} when the algorithm was taking into account one of
the nodes $n_1,...,n_k$.
Conversely, every node $n_1,...,n_k$ is also responsible for
one of the arcs $a_1,...,a_k$. Indeed, if a node $n_i$ causes
the selection of an arc $\hat a \notin \{a_1,...,a_k \}$, then
we are left with the $k$ arcs which cannot all be chosen
by $k-1$ nodes.\\
Let us consider the node $n_1$. Without loss of generality
assume that it is responsible for the selection of the arc $a_1$
with weight $d_1$ and linking it to the node $n_2$.
According to the previous results,
$n_2$ can not be  responsible for the choice of $a_1$. Let $a_2$
be the arc selected because of $n_2$ with weight $d_2$ and
connecting it to $n_3$. Observe that necessarily $d_2\leq d_1$.
We may repeat this process till the node $n_{k-1}$. Hence,
we obtain that every node $n_i$ is connected to $n_{i+1}$
by the arc $a_{i}$ whose cost is $d_{i}\leq d_{i-1} $, for 
$i=2,...,k-1$. Finally consider $n_k$. It must be responsible for
$a_k$ which has to connect it to $n_1$ with cost $d_k\leq d_{k-1}$.
Since $d_k$ is incident to $n_1$ it holds that $d_1 \leq d_k$
Therefore $d_1\leq d_k \leq d_{k-1} ...\leq d_2 \leq d_1$ and
we have the assertion of the second property.\\
About the third property, the upper bound $N$ follows from the
consideration that every node causes the choice
of at most a new arc.
In step {\tt 2c} of the algorithm, it may happen at most
$\lfloor N/2 \rfloor$ times that we are forces to pick up an arc
which is already in $A$.
So we have at least $N-\lceil N/2 \rceil = \lfloor N/2 \rfloor$ arcs
$\hfill \square$

\bibliography{ifac}

\begin{thebibliography}{10}

\bibitem{blatt96}
M.~Blatt, S.~Wiseman, and E.~Domany.
\newblock Super-paramagnetic clustering of data.
\newblock {\em Physical Review Letters}, 76:3251, 1996.

\bibitem{rava02}
E.~Ravasz, A.~L.Somera, D.~A. Mongru, Z.~N. Oltvai, and A.~L. Barabási.
\newblock Hierarchical organization of modularity in metabolic networks.
\newblock {\em Science}, 297:1551, 2002.

\bibitem{palmer84}
R.~G. Palmer, D.~L. Stein, E.~Abrahams, and P.~W. Anderson.
\newblock Models of hierarchically constrained dynamics for glassy relaxation.
\newblock {\em Phys. Rev. Lett.}, 53(10):958--961, Sep 1984.

\bibitem{mant99}
R.~N. Mantegna.
\newblock Hierarchical structure in financial markets.
\newblock {\em Eur. Phys. J. B}, 11:193--197, 1999.

\bibitem{mardia79}
K.~V. Mardia, J.~T. Kent, and J.~Bibby.
\newblock {\em Multivariate Analysis}.
\newblock Academic Press, London, UK, 1979.

\bibitem{anderson73}
M.~Anderberg.
\newblock {\em Cluster Analysis for Applications}.
\newblock Academic Press, New York, 1973.

\bibitem{eisen1998clu}
M.~B. Eisen, P.~T. Spellman, P.~O. Brown, and D.~Botstein.
\newblock Cluster analysis and display of genome-wide expression patterns.
\newblock {\em Proc Natl Acad Sci U S A}, 95(25):14863--8, 1998.

\bibitem{tum07}
M.~Tumminello, C.~Coronnello, F.~Lillo, S.~Miccich\'e, and R.~N. Mantegna.
\newblock Spanning trees and bootstrap reliability estimation in
  correlation-based networks.
\newblock {\em Int. J. of Bifurcations \& Chaos}, 17:2319--2329, 2007.

\bibitem{mant95}
R.~N. Mantegna and H.~E. Stanley.
\newblock Scaling behaviour in the dynamics of an economic index.
\newblock {\em Nature}, 376:46--49, 1995.

\bibitem{stan99a}
P.~Gopikrishnan, V.~Plerou, L.~A.~N. Amaral, M.~Meyer, and H.~E. Stanley.
\newblock Scaling of the distributions of fluctuations of financial market
  indices.
\newblock {\em Phys. Rev. E}, 60:5305--5316, 1999.

\bibitem{nrm07}
M.~J. Naylora, L.~C. Roseb, and B.~J. Moyle.
\newblock Topology of foreign exchange markets using hierarchical structure
  methods.
\newblock {\em Physica A}, 382:199--–208, 2007.

\bibitem{ljung99}
Lennart Ljung.
\newblock {\em System identification: theory for the user (2nd Ed.)}.
\newblock Prentice-Hall, Inc., Upper Saddle River, NJ, USA, 1999.

\bibitem{ksh00}
T.~Kailath, A.~H. Sayed, and B.~Hassibi.
\newblock {\em Linear Estimation}.
\newblock Prentice Hall, Upper Saddle River, New Jersey, 2000.

\bibitem{gra69}
Clive W.~J. Granger.
\newblock Investigating causal relations by econometric models and
  cross-spectral methods.
\newblock {\em Econometrica}, 37:424--438, 1969.

\bibitem{kail01}
A.~H. Sayed and T.~Kailath.
\newblock A survey of spectral factorization methods.
\newblock {\em Numerical Linear Algebra with Applications}, 8:467--469, 2001.

\bibitem{shir95}
A.~N. Shiryaev.
\newblock {\em Probability}.
\newblock Springer-Verlag, New York, 1995.

\end{thebibliography}

\end{document}